# IMPROVING TCP PERFORMANCE OVER WIRELESS NETWORK WITH FREQUENT DISCONNECTIONS

Purvang Dalal[1], Nikhil Kothari[1] and K. S. Dasgupta[2]

[1] Department of Electronics and Communication, D.D.University, Nadiad, Gujarat.
`pur_dalal@yahoo.com, nil_kothari@yahoo.co.in`
[2] IIST, Trivendrum, INDIA
`ksd@iist.ac.in`

## ABSTRACT

*Presented in this paper is the solution to the problem that arises when the TCP/IP protocol suite is used to provide Internet connectivity through mobile terminals over emerging 802.11 wireless links. Taking into consideration the strong drive towards wireless Internet access through mobile terminals, the problem of frequent disconnections causing serial timeouts is examined and analyzed, with the help of extensive simulations. After a detailed review of wireless link loss recovery mechanism and identification of related problems, a new scheme with modifications at link layer and transport layer is proposed. The proposed modifications which depend on interaction between two layers (i) reduce the idle time before transmission at TCP by preventing timeout occurrences and (ii) decouple the congestion control from recovery of the losses due to link failure. Results of simulation based experiments demonstrate considerable performance improvement with the proposed modifications over the conventional TCP, when a wireless sender is experiencing frequent link failures.*

## 1. INTRODUCTION

Due to advancement of wireless technology and the proliferation of 802.11[1] based hand-held wireless terminals, recent years have witnessed an ever-increasing popularity of wireless networks, ranging from Wireless Local Area Networks (WLANs) to Mobile Adhoc Networks (MANETs) [2-4]. In integrated WAN + LAN + 3G cellular systems, data and multimedia communications are carried end to end over the existing Internet infrastructure. In WLANs (Wi-Fi technology using 802.11) mobile hosts communicate with an Access Point (AP) or a Base Station (BS) that is connected to the wired networks. Visibly, only one hop wireless link is needed for communications between a Mobile Host (MH) and a stationary Fixed Host (FH) in wired networks. Most of the data traffic over the WLAN is TCP traffic, including traffic generated by web accesses, e-mails, bulk data transfers, remote terminals, etc [3-6]. However, TCP/IP needs to meet the challenges introduced by the wireless portion of the network in order to make it compatible with the wired network for providing efficient Internet services.

When TCP packet loss occurs at a congested link, recovery mechanism is triggered at sender TCP either on arrival of Duplicate Acknowledgements (*dupacks*) or expiration of sender's Retransmission timer [7]. To relieve the link congestion, TCP transmits at a lower rate by shrinking its transmission window or congestion window (*cwnd*). Thus TCP's loss recovery mechanism is unconditionally coupled with congestion control mechanism. Such TCP behaviour works fairly well in wired networks, where packet losses are almost caused by link congestion; and packet loss due to the random Bit Error Rate (BER) is either negligible or not exceeding one packet loss per *cwnd* [8, 9]. When TCP works over wireless environments, several known problems affect its performance because of the differences in terms of bandwidth, propagation delay, and link reliability [9-12]. TCP connections encounter other types of delays and packet losses than those are unrelated to congestion [8]. First, packets may be lost due to high BER and channel asymmetries (both in the presence and in the absence of user mobility). Second, communication may pause during hand-off between cells. Third, packets may be lost, while a mobile host is out of reach from other transceivers.

In a wired-cum-wireless data connection, the IEEE 802.11 wireless link from the BS to the MH is frequently getting disconnected due to excessive mobility or power crises [13]. This type of disconnections typically last longer than a Retransmission Timeout (*RTO*) at TCP, while potentially being shorter than the lifetime of a TCP connection. A serious problem caused by this temporal disconnection is of serial timeouts at the TCP sender wherein multiple consecutive retransmissions of the same packet are attempted for the disconnected destination [14, 15]. Since the *RTO* interval is doubled with each unsuccessful retransmission (until it reaches 64s [15]), several consecutive failures can lead to inactivity lasting several minutes. This increases idle time at sender TCP even when the link is re-established. The packet flow also reduces because of the drastic reduction in *cwnd*, which introduces additional delay for raising *cwnd* upto a reasonable value so as to utilize network capacity. Thus serial timeouts can be more harmful to overall throughput, compared to losses due to bit errors [16]. Hence this kind of timeouts should be avoided.

In this paper, the TCP performance issues over a network suffering from frequent link failures are studied through simulations under *ns-2*. After identifying the problems, a modified TCP scheme, which reduces the impact of idle time, is proposed. This proposed scheme utilizes channel immediately after expiration of disconnection. Moreover, the scheme retains the *cwnd* at the sender once the link is re-established. The superiority of the proposed scheme is demonstrated by comparing the end to end performance of the same with original scheme. Rest of the paper is organized as follows: In Section 2, the previous work in improving the TCP performance for wireless losses related to link disconnections is reviewed. The authors also highlight the important issues, which have remained unresolved. Section 3 defines the problem based on study of the TCP loss recovery and related performance issues, using a trace-file. In Section 4, modifications at Medium Access Control (MAC) layer and TCP layer respectively for improving TCP performance are presented. In section 5, the schemes with the help of simulations under different network conditions are evaluated. The paper concludes in Section 6 with a summary of the results and highlight of future work.

## 2. RELATED WORK

A lot of research to mitigate the performance problem of TCP, related to the wireless mobile environment is happening and as a result, several versions have emerged [16-21]. Earlier attempts to address this problem can be broadly divided into two groups. The first group does not attempt to change or modify the TCP protocol, instead exploits methods such as injecting, removing or delaying TCP packets based on a superior understanding about what is happening at the link layer. These mechanisms provide solutions at the Link Layer (LL), by hiding the deficiencies of a wireless channel from TCP, referred as TCP *unaware*. On the other hand, the second group endeavours for modifying the behaviour of TCP. These mechanisms are referred as TCP *aware*. In this section, both TCP unaware solutions and TCP aware solutions are discussed.

### 2.1. TCP *un-aware* Solutions

LL protocol running on top of the physical layer is more adaptable to link characteristics [22] than higher-level protocols and it is much faster as it has immediate knowledge of dropped frames. Alleviating the inefficiencies of the wireless medium at the LL provides the transport layer protocol with a dependable communication channel, similar in characteristics to a wired channel. This is realized using an Automatic Repeat reQuest (ARQ) scheme and/or by means of Forward Error Correction (FEC) [23]. ARQ in presence of high error rates can lead to a large volume of retransmissions and can even cause a complete black-out in the connection. On the other hand, FEC is not very well suited for channels with bandwidth constraints common in wireless environments. It also increases power requirement and introduces computation delays for each packet. The '*Delayed Dupacks*' scheme [24] looks at the same problem but in the context of a reliable link layer protocol which acknowledges each packet and performs fast retransmission. TULIP (*Transport Unaware Link Improvement Layer*) [24] also attempts to recover from retransmission losses before TCP coarse-grain timeouts occur. These schemes, in absence of interaction between TCP and MAC protocols lead to futile TCP retransmission [23], causing wastage of scarce resources and performance problems on end to end basis. In order to overcome limitations of the above mechanisms, TCP has to be made aware of what is happening at link layer for making it adaptable to the link characteristics.

## 2.2. TCP *aware* Solutions

Here the approach is to divide the end-to-end connection between a FH and a MH into two separate connections at the wired-wireless border (e.g., BS). Major problem with split connection in approaches like I-TCP [25] and M-TCP [26] is hand-off latency due to mobility of MH between BS besides violation of the end-to-end semantic of TCP [27]. Additionally, these approaches are vulnerable to scalability issues due to requirement of per-flow support from the BS or modification in TCP at the FH [27]. These schemes demand for modifications at intermediate nodes; which also create practical issues at the time of deployment in existing networks. Taking into consideration these limitations, the authors have focused on approaches with modifications only at the MH. In *Freeze TCP* [28], upon receipt of an indication of disconnection from lower layer, TCP at the MH sends a *Zero Window Advertisement* (ZWA) to the FH, to pause transmission. Upon reconnection, a fast retransmit is adopted to restart transmission. The main drawback of *Freeze TCP* is dependency on the network layer to predict future disconnections [27]. The approaches like M-TCP and Freeze TCP may also trigger bursty TCP transmissions and affect the competing traffic [27]. To our knowledge, certain issues as listed below have remained unresolved.

1) Majority of TCP schemes largely rely on signals from end systems or an intermediate node for resuming transmissions after link re-establishment [27]. In presence of network latency, the above schemes would not be able to perform well as the triggering of transmission with negligible delay is unlikely to happen.
2) Futile link layer re-transmissions for loss recovery over a wireless link further delay transmission of subsequent TCP packets from its buffer. This results into false Round Trip Time (*RTT*) estimation leading to an increased *RTO* interval. A longer *RTO* interval subsequently prevents TCP from regaining the available network bandwidth, sooner.

Therefore, there is a need for a solution which can improve TCP performance in wireless environment suffering from frequent disconnections, without sacrificing end to end TCP semantics. In the next section, the impact of link failures on the performance of TCP is examined, in order to define the problem, related to the inherent weaknesses of TCP.

## 3. PROBLEM DEFINITION

Figure 1 shows a typical trace-file output obtained from simulation in a WLAN having link failure. The authors analyzed the trace-file to understand the problem of TCP's inability to attempt transmissions immediately after link restoration, as pointed out in the previous section. As seen in Fig.1, the wireless link remained disconnected from 27.75 sec to 27.94 sec. TCP packet 10119 was transmitted at 27.719 sec. All previous TCP packets belonging to the current *cwnd* were successfully acknowledged. TCP packet 10119 was transmitted by MAC at 27.75 sec, during the link disconnection. After failure of all consecutive MAC retransmissions attempts (decided by *Retry Limit – RL*, usually 6 [29]), the MAC layer discards the TCP packet and takes up the next TCP packet from an *InterFace Queue (IFQ)*. This process is repeated for all outstanding TCP packets in the *IFQ*. After dropping the last TCP packet 10128, MAC layer is required to wait till another packet is queued from TCP. TCP resumes transmission (in this case retransmission after timeout, *RTO* of 240 ms) at 27.95 sec. It was observed that the TCP transmission is unnecessarily delayed by 18.5 ms, even after the earlier restoration of the link at 27.94 sec.

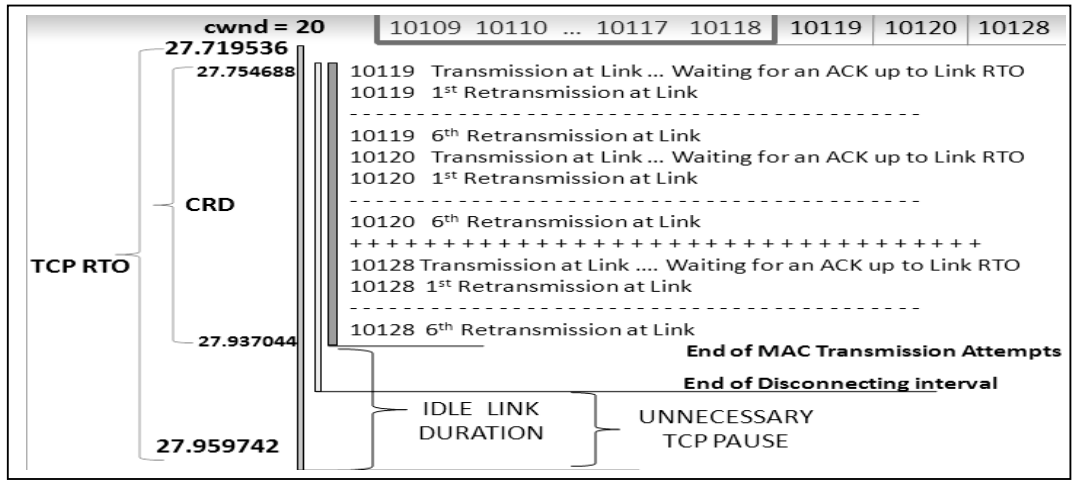

Figure. 1 Trace file based analysis for MAC loss recovery (*With Default RL = 6*)

The delayed loss recovery also effects the self clocking growth of *cwnd*. It is noticed that once the successful retransmission is attempted, TCP is able to restore cwnd to its value prior to link disconnection after 1.98 sec. Following observations related to performance degradation are made based on the above study.

1) TCP is unable to utilize link, immediately after its restoration.
2) The network capacity is underutilized in the initial period, immediately after resuming of transmissions.

These problems are elaborated with the help of an illustration in Figure 2; wherein the disconnection interval is from $T_0$ to $T_4$. After transmission of last packet from current *cwnd* at time $T_1$, TCP has to wait either for arrival of *acks* or expiration of *RTO*.

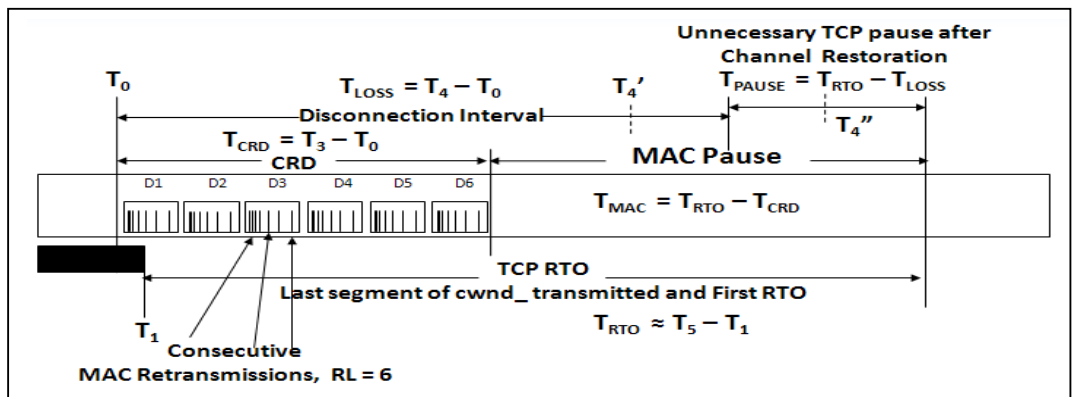

Figure. 2 Graphical illustration of $T_{CRD}$ and $T_{PAUSE}$

During *RTO* period MAC layer attempts transmission of all outstanding TCP packets from *IFQ*. This duration is referred as *Consecutive Retry Duration ($T_{CRD}$)*. After $T_{CRD}$, there is no transmission at MAC layer until another transmission attempt at TCP, resulting into idle time at MAC layer. Under this condition, retransmission at TCP resumes only after expiration of *RTO* and introduces idle time at TCP, following restoration of the link as shown in Figure 2. Idle time at TCP after restoration of the link can be mathematically related with *RTO* and the disconnection interval ($T_{LOSS}$) as shown below.

$$T_{PAUSE} = RTO - T_{LOSS} \quad \dots\dots\dots\dots\dots\dots\dots\dots\dots\dots\dots\dots\dots\dots\dots\dots\dots\dots\dots\dots\dots \quad (1)$$

Eq.1 holds true whenever $T_{LOSS} > T_{CRD}$. It is apparent that $T_{PAUSE} \approx 0$, for $T_{LOSS} \leq T_{CRD}$, as transmission at MAC after link restoration brings acks for TCP and triggers transmission before *RTO*. Note that for $T_{LOSS} > T_{CRD}$, TCP retransmission occurs after TCP timeout, with considerable delay following the restoration of the channel i.e. $T_{PAUSE}$. Thus, TCP performance degrades due to non-utilization of the channel during $T_{PAUSE}$. Here, it may be noted that with increase in $T_{LOSS}$ within *RTO*, $T_{PAUSE}$ reduces.

Furthermore, the packet loss is interpreted as a network congestion and the retransmission follows the conventional algorithm with reduced flow, i.e. halving the *ssthresh* (slow start threshold); slow restart (*cwnd* decreases to 1). This reduction in transmission flow at TCP for wireless losses is inappropriate. Suppose the value of *cwnd* prior to link disconnection is *cwnd(n)* and the value using which TCP resumes its transmission is *cwnd(n+1)*, then time required to raise cwnd from *cwnd(n)* to *cwnd(n+1)* is referred here as TCP response time ($T_{RES}$), during which TCP under-utilizes the network capacity. This adversely affects on end to end TCP performance. As growth in *cwnd* is *RTT* dependent, over a network with larger *RTT*, longer is the $T_{RES}$ and similar will be the impact on its performance.

In next section, the approaches are proposed to restart transmissions immediately after link re-establishment and to improve utilization of network capacity further by restoring *cwnd* earlier at the same instant.

## 4. PROPOSED MODIFICATIONS

As discussed earlier, the authors adopt an approach, which primarily follows the scheme used in TCP *unaware* solutions. This approach improves TCP's performance with efficient loss recovery at MAC layer. This improves TCP's performance by a) hiding most of the wireless losses from TCP and b) reducing transmission pause at TCP. The approach is further extended with a TCP *aware* solution to get improved network utilization by making it more adaptable to the link characteristics. This enables TCP to decouple its congestion control algorithm while attempting recovery of the packets lost during disconnection. Thus the pair of approach is a combination of TCP *aware* and *unaware* solutions.

### 4.1. Approach A – MAC layer modifications

$T_{PAUSE}$ can be reduced by extending $T_{CRD}$ over $T_{LOSS}$. Mathematically $T_{CRD}$ is represented as

$$T_{CRD} = \sum_{n=1}^{s} \left( [RL+1] T_{LINK} + \sum_{k=0}^{RL} T_{BACKOFF}(k) \right) \quad \dots\dots\dots\dots\dots\dots\dots\dots \quad (2)$$

Here, s is the number of outstanding TCP packets in an *IFQ*. The first term of the sum correspond to the constant part of the duration necessary for the transmission of a segment and its *RL* retransmissions. $T_{LINK}$ represents two way transmission delay over a link and processing delay at nodes connected to it. $T_{LINK}$ is thus a time including all the different constant durations of the MAC 802.11 protocol (excluding those related to the *RTS/CTS* mechanism) [29].

The second term of $T_{CRD}$ corresponds to the sum of the back-off duration for each segment and its successive retransmissions according to the index *k* of the retransmission, decided by the MAC layer parameter *Contention Window (CW)* [29]. As per Eq. 2, $T_{CRD}$ can be increased by tuning different TCP and MAC parameters like (a) enlarged *cwnd* at the TCP layer, (b) a higher *RL* at the MAC layer, (c) a larger *CW* at the MAC layer and (d) an increased $T_{LINK}$ by reducing data-rate. Enlarging *cwnd* is not desirable as it can lead to large number of TCP retransmissions in the case of network congestion. *CW* increase at the MAC layer is unsuited because the duration of this window is also used for contention resolution when several stations are competing to access the same channel [29].

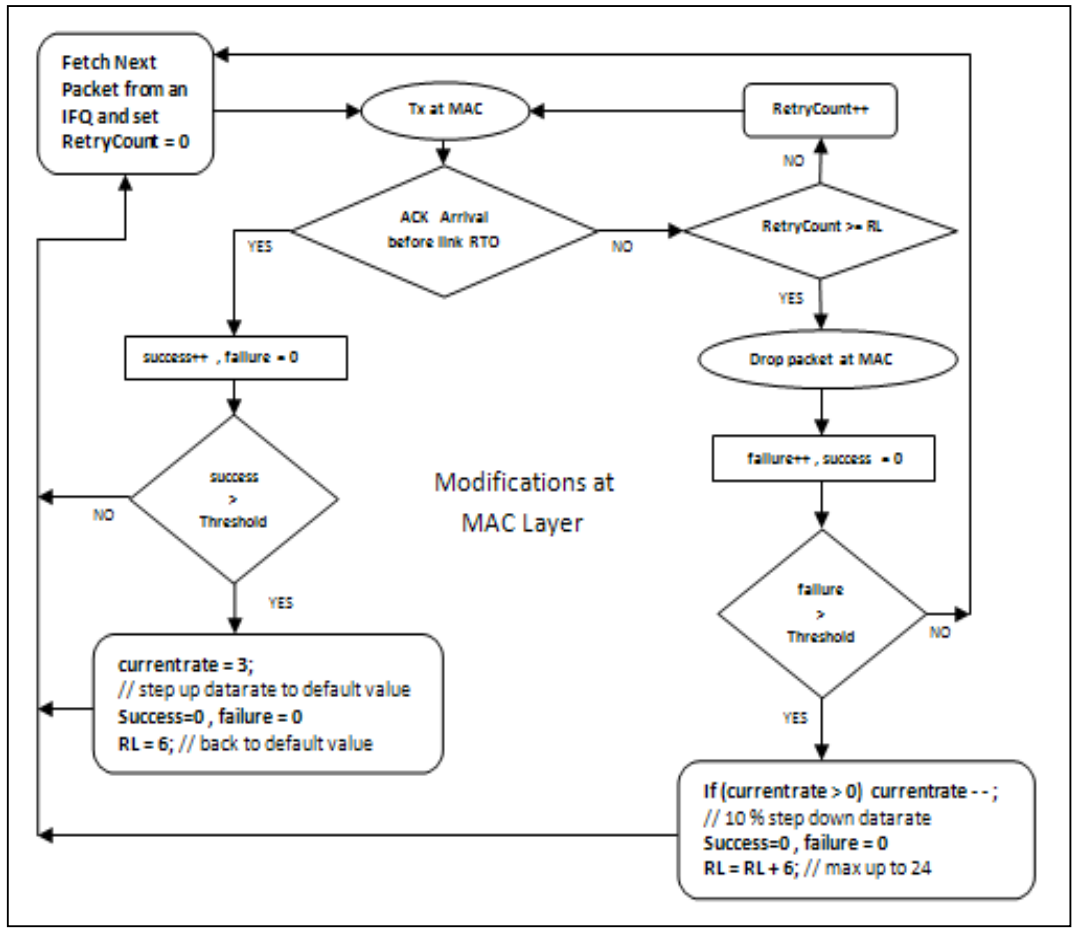

Figure. 3 Flowchart for MAC Layer modifications

Considering all above, an approach is proposed that increases $T_{CRD}$ with the help of increase in (i) $T_{LINK}$ and (ii) *RL*. The authors have considered increase in *RL* for extending $T_{CRD}$, beyond disconnection. This improves MAC loss recovery with more retransmission attempts over a link. The increase of the *RL* is linear and progressive (the default value of 6 is successively added), based on the successive transmission failures detected at MAC layer. Under lossy conditions decreasing data-rate is also anticipated as per *ARF (Auto Rate Fallback)* scheme, which not only reduces the probability of transmission losses but also increases $T_{LINK}$. Hence, in addition to increase in *RL*, the authors also considered reduction in data-rate during each step; with maximum reduction up to 33% in 3 consecutive steps (refer to the flowchart in Figure 3). Thus, $T_{CRD}$ is increased with the help of rise in *RL* and reduced data-rate.

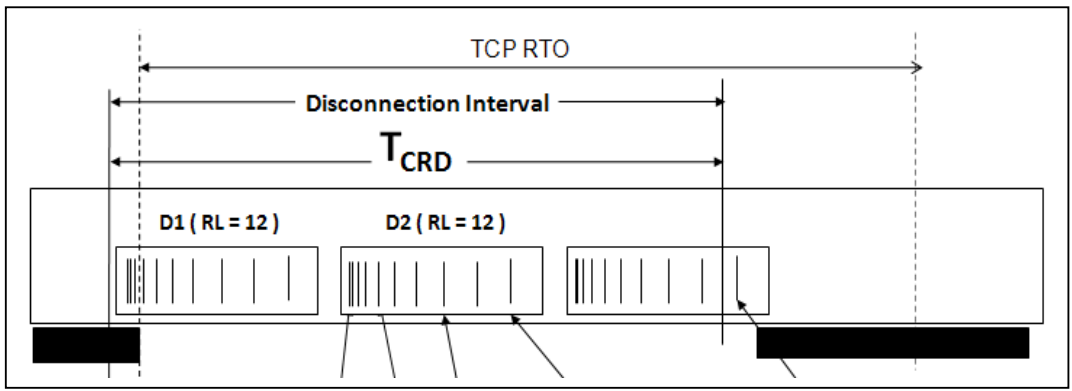

Figure. 4 Illustration of improved MAC loss recovery with extended $T_{CRD}$

Figure 4 graphically presents the approach proposed to reduce idle time at TCP after restoration of channel. As shown, Packets $D_1$ and $D_2$ are dropped after attempting 12 MAC retransmissions. With this approach, $T_{CRD}$ is higher than $T_{LOSS}$, implying that losses do not cause a TCP retransmission after timeout (*RTO*). As soon as the channel is restored, the latest TCP segment (here packet $D_3$) is successfully retransmitted by the MAC layer and it reaches the receiver. Afterwards, the sender can continue its transmissions until its TCP window is emptied. The segment not received is finally retransmitted by the sender after the reception of *3 dupacks*. The time wasted before this retransmission is therefore relatively weak compared to $T_{PAUSE}$, observed earlier. Hence, the increase of the $T_{CRD}$ provides a recovery mechanism faster than a TCP retransmission after *RTO*. Improved MAC loss recovery with the help of delayed/increased MAC retransmissions facilitates TCP by a) minimizing $T_{PAUSE}$ and b) converting costlier timeouts into fast re-transmissions. On the other hand, it leads to excessive MAC retransmissions. However, the fairness with other traffics on the same channel is preserved due to the 802.11 back-off algorithm.

It should be noted that the *RL* and data-rate are restored back to default value on the arrival of the MAC acknowledgment for the retransmitted segment. Therefore, this modification does not adversely affect the overall performance under idle channel conditions.

### 4.2. Approach B – MAC and TCP layer modifications

In previous approach, the increase in $T_{CRD}$ is based on detection of successive MAC retransmission failures. Hence, even with extended loss recovery at MAC, few TCP packets are likely to be dropped at MAC. As described earlier, TCP triggers recovery for such losses using fast retransmit, earlier than recovery after timeout. Unfortunately, the earlier transmission is attempted at reduced rate due to halving of *cwnd*, as per the conventional TCP behaviour. This prevents TCP to utilize network to its capacity for the duration $T_{RES}$. Hence in addition to approach A, an attempt to decouple congestion control and loss recovery for the losses occurring due to link failures, is considered. As per this approach B, TCP avoids flow control while attempting recovery of the packets lost during disconnection, based on the indication from link failure. Algorithm implemented at MAC and TCP layer is as shown in Figure 5. TCP Newreno is adopted as a standard baseline protocol for attempting modifications.

Wireless link failure is detected at MAC layer and subsequently informed to TCP with the help of a parameter *link_loss*, which is accessible at both layers. In case of a failure of MAC loss recovery using approach A, for any of the TCP packet from current *cwnd*, the parameter is set to 1 at MAC layer. The algorithms at TCP and MAC layers, manipulate the parameter as follows. (i) *link_loss* is reset to default value 0 at TCP. (ii) Whenever a TCP packet is dropped, MAC layer sets *link_loss* to 1. Based on the high status of *link_loss*, TCP attempts loss recovery without halving *cwnd*. As a result, TCP not only resumes transmission immediately after link restoration but also improves network utilization. After attempting loss recovery using the modified scheme, TCP reverts back to its default state by resetting *link_loss* to 0. This ensures triggering of conventional congestion control for other packet losses detected then after. It should be noted that the MAC layer does not set *link_loss* to 1 for the congestion induced losses as they are not visible at link layer.

```
AT MAC Layer
For transmission of each segment from the
current cwnd of n segments.

link_loss = 0;
for(i=0;i<n;i++)
        RetryCount = 0;
        while (1)
        {
                transmit();
                If(ack_rcvd())  break;
                RetryCount++;
                If(RetryCount >=RL)
                {
                        link_loss = 1;
                        break;
                }
        }
}
```

```
AT TCP Layer
For loss recovery of any packet loss detected
using 3 DupAck

if (link_loss == 1)
{
        link_loss = 0;
        // do not modify cwnd/ssthresh
}
else
{
        // trigger TCP Congestion Control
        // change cwnd / ssthresh
        .......
        ............
        cwnd = cwnd / 2;
}
```

Figure. 5 pseudo codes for MAC and TCP Layer modifications

## 5. SIMULATION RESULTS & ANALYSIS

In this section, the impact of link disconnections on $T_{PAUSE}$ is examined with the help of simulation based experiments. Subsequently, the improvement in performance of TCP by incorporating the changes proposed in the approaches discussed earlier is also evaluated. It may be noted that the similar approach is used with any of the TCP variant with the modification in its basic loss recovery mechanism. However in this paper TCP Newreno is used considering it as the trusted baseline algorithm by the entire group of researchers.

### 5.1. Simulation Environment

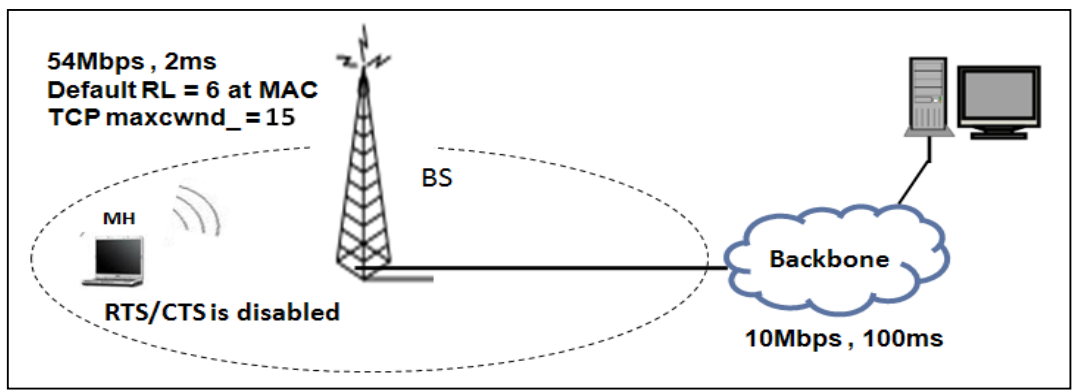

Figure. 6 Wireless Network Scenario under *ns-2*

Extensive simulations over a controlled wireless network, using *ns-2* [30] were conducted. All simulations discussed in this paper are confined to the wireless networks with IEEE 802.11 *Distributed Coordination Function (DCF)* [31] compliant wireless links. The typical simulation scenario is as shown in figure 6, for all experiments performed and analysed. The authors have considered only first link as a wireless link and all losses are introduced on the same link. The backbone of a hybrid network is represented by a wired portion with 10Mbps of datarate and 100ms of delay.

### 5.2. Simulation Parameters

The typical simulation scenario and other simulation parameters used in majority of experiments are illustrated in Figure 6, above. The authors have considered variations in the parameters, over the first wireless link. Wireless losses are introduced over a link between BS and MH. Desired disconnection interval is ensured with the help of controlled mobility of MH, in and out of the coverage area of BS. The interval is kept long enough to allow MAC layer to exploit its loss recovery for all TCP packets from an *IFQ*. During simulations default MAC parameters were used; i.e. $RL = 6$ and maximum allowable data-rate. FTP traffic between MH and FH is generated from 5 sec to 15 sec of duration whereas the disconnection is introduced at 10.2 sec. During experiments 12 TCP packets from current *cwnd* were unacknowledged, just before disconnection. The authors have considered $T_{LOSS}$ above 40 ms, where $T_{LOSS} > T_{CRD}$. This condition is chosen to understand impact of increase in $T_{CRD}$ on $T_{PAUSE}$, and also on end to end TCP performance.

### 5.3. Simulation Results

Simulations were conducted initially with an objective to understand impact of $T_{LOSS}$ on $T_{PAUSE}$. In initial set of experiments $T_{LOSS}$ is increased in step from 0.2 *RTT* to *RTT*. $T_{PAUSE}$ under different $T_{LOSS}$ durations is measured and plotted as shown in Fig. 7. Since the *RTO* interval is fixed, $T_{PAUSE}$ reduces linearly with increase in $T_{LOSS}$ according to the Eq.1, explained earlier.

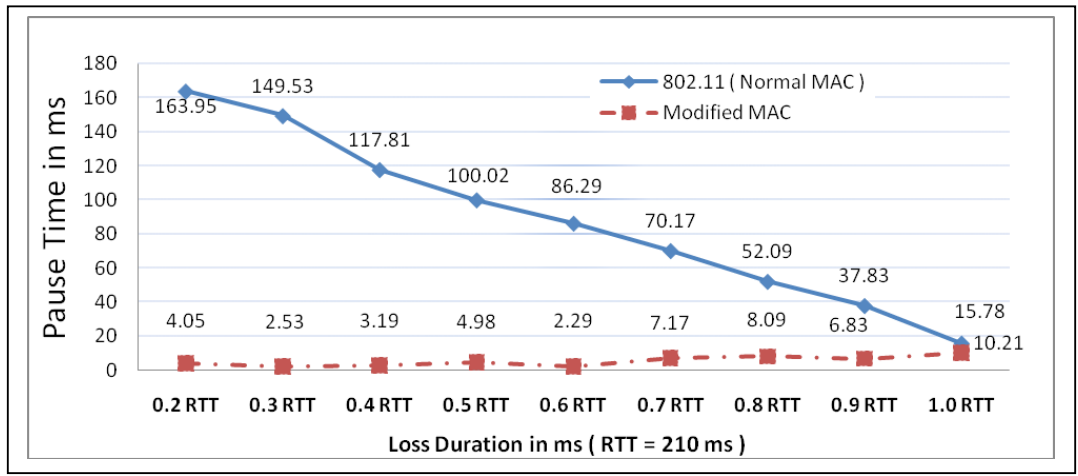

Figure. 7 $T_{PAUSE}$ Vs $T_{LOSS}$
*($T_{LOSS}$ <= RTO and short FTP with Single Disconnection)*

Simulations were repeated with the modified MAC layer as per approach A, which modifies the value of *RL* and data-rate dynamically depending upon the number of consecutive failures experienced. This causes a larger $T_{CRD}$ resulting into a significant reduction in $T_{PAUSE}$. It was observed that the modifications produced $T_{CRD}$ larger than $T_{LOSS}$ in all the cases of loss durations of Figure 7. A Retransmission attempted by MAC layer during $T_{CRD}$, particularly continuing beyond $T_{LOSS}$, successfully reaches the receiver. This successful MAC layer retransmission may produce either *ack* or *dupacks* at TCP. In case of arrival of *ack*, the transport layer is protected from the link layer problems as it is *unaware*. The arrival of dupacks triggers fast recovery at the transport layer. In either condition the transport layer performance is improved due to far-reaching retransmission attempts at MAC. Further the TCP is also protected from performance degradation due to an unnecessary *RTO*. In Figure 7 the $T_{PAUSE}$ with MAC layer modifications is observed to be marginally higher than 0, because of inevitable transmission delay at MAC layer.

In order to evaluate the impact of the modification on end to end performance resulting from reduction in $T_{PAUSE}$, TCP throughput with different MAC implementations over the same network scenario is measured. Figure 8 quantifies the performance improvement at TCP, with the modified MAC layer. Performance gain is dependent on the amount of reduction in $T_{PAUSE}$ (refer to figure 7), which is continuously decreasing, as shown in Figure 8. This is owing to the fact that the MAC layer modifications have more opportunity to play role with the higher value of $T_{PAUSE}$. It should be noted that this improvement is mainly because of the efficient MAC loss recovery and prevention of costlier TCP timeout during single loss duration. The relative performance of the modified scheme will improve further in presence of frequent disconnections as it will protect TCP from unnecessary timeouts over and over again with the help of elimination of $T_{PAUSE}$. In practical situations, disconnections in a mobile wireless networks are frequent with variations in time. Thus, the proposed modifications would also act frequently and prevent more TCP timeouts depending on the number of disconnections.

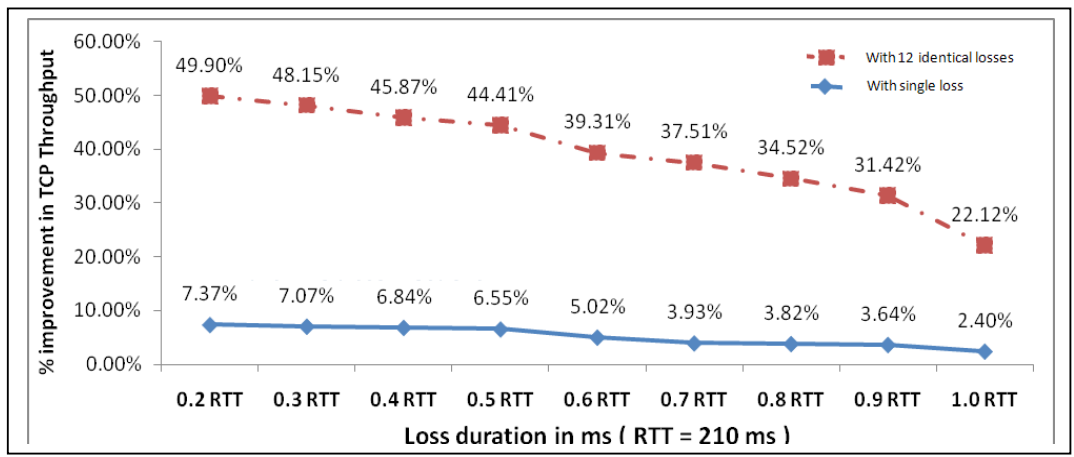

Figure. 8 Comparison in % improvement in TCP Throughput Vs $T_{LOSS}$
*($T_{LOSS}$ <= RTO and short FTP)*

The authors also performed simulations to confirm effectiveness of the proposed scheme in presence of above mentioned network environment. In this case, a FTP data was transferred for 250 ms of the simulation period having 12 identical disconnections. Reduction in $T_{PAUSE}$ for each of these 12 disconnections leads to significant performance gain at TCP as shown in Figure 8. However, $T_{PAUSE}$ reduces with increase in $T_{LOSS}$, which in turn affects the relative performance gain.

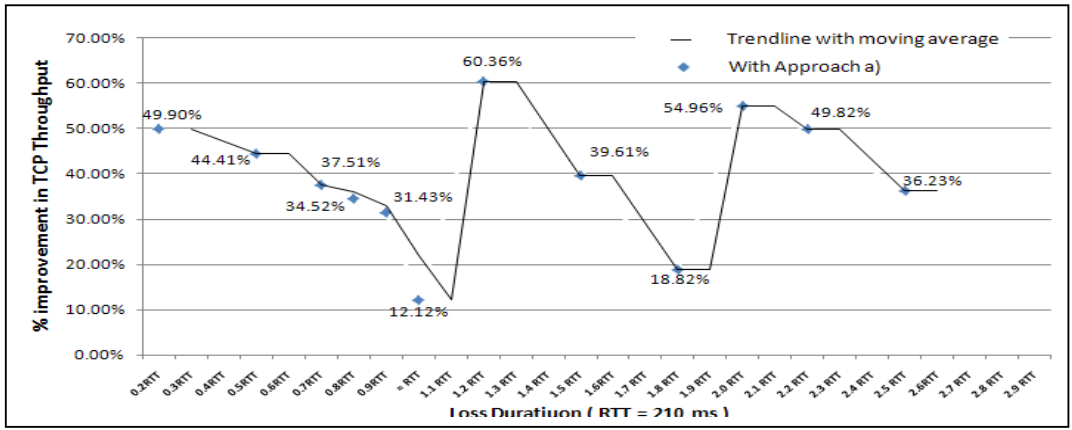

Figure. 9 Comparison in % improvement in TCP Throughput Vs $T_{LOSS}$
($T_{LOSS}$ > RTO and Long FTP with 12 identical disconnections)

If the disconnection continues beyond the timeout interval ($T_{LOSS}$ > $RTO$), retransmission timeout is inevitable. Further the retransmission attempted after $RTO$ is also lost due to unavailability of the channel and $RTO$ interval increases exponentially leading to a larger $T_{PAUSE}$ in conventional TCP. Note that a larger $T_{PAUSE}$ in conventional TCP, benefits the relative performance improvement in the modified TCP, which is able to reduce $T_{PAUSE}$ effectively. However, increase in $T_{LOSS}$ between consecutive $RTO$ intervals has an impact on $T_{PAUSE}$ similar to that observed in case of single loss duration. $T_{PAUSE}$ was found to be continuously decreasing in this case also. Since the performance gain at TCP is proportional to the reduction in $T_{PAUSE}$, improvement also reduces with the modified scheme as shown in Figure 9. Figure demonstrates such repetitive behaviour in the range of $RTO$ (following exponential back-off effect).

In order to substantiate the performance improvement in presence of the modifications proposed in this paper, the number of TCP packets dropped at MAC layer in either condition is compared in Table 1. As seen in the table the default MAC behaviour is suffering from large number of packet drops during disconnection, because of i) smaller $RL$ and maximum data-rate ii) exponential back-off in $RTO$, causing futile TCP transmissions. On the other end a larger $RL$ with reduced data-rate benefit the loss recovery at link layer in presence of the proposed modifications. In the modified scheme, extended MAC loss recovery for each packet reduces the number of MAC drops. It is obvious that reduction in packet drops at MAC layer reflects into requirement of retransmission attempts at TCP for loss recovery. Reduction in number of TCP retransmissions makes TCP more efficient by allowing it to probe and utilize the network capacity further.

Table 1. TCP packet drops at MAC layer.

| Loss Duration in RTT (RTT = 210 ms) | TCP packets dropped at MAC (total 12 identical losses) | |
|---|---|---|
| | **Normal MAC** | **Normal MAC** |
| 0.5 RTT | 22 | 12 |
| 1.0 RTT | 37 | 17 |
| 1.5 RTT | 54 | 24 |
| 2.5 RTT | 141 | 40 |
| 4.0 RTT | 179 | 82 |
| 5.0 RTT | 176 | 91 |
| 6.0 RTT | 173 | 102 |

In presence of modification, MAC layer's continuous retransmission attempts for the same TCP packet may bring *ack* for TCP; hiding link layer discrepancies. However TCP may receive *dupacks* in response to the packet successfully reaching the receiver after link re-establishment; caused by a prior loss during the disconnection. This invokes fast recovery and halving of *cwnd* [11, 20] in absence of congestion. Hence, the network capacity remains under-utilized till *cwnd* is raised to the value prior to the disconnection. This period of $T_{RES}$ was observed close to 9 times the value of *RTT* in the experiments.

Observations to illustrate impact of modifications on $T_{PAUSE}$ and $T_{RES}$, with different approaches under evaluation are presented in Figure 10. As shown in the Figure, the modified scheme with only approach A has *cwnd* larger than the conventional TCP. The modified scheme shows a higher *cwnd* because of timeout avoidance. However it is unable to prevent *cwnd* halving at 11.002 sec for the reason discussed above. Consequently the modified scheme with only approach A spends considerable time in regaining the prior value of *cwnd* ($T_{RES}$ = 1.89 sec). Extension of approach A with approach B enables the modified TCP to retain *cwnd* based on the signal from MAC layer. As a result network utilization is improved, immediately after link restoration.

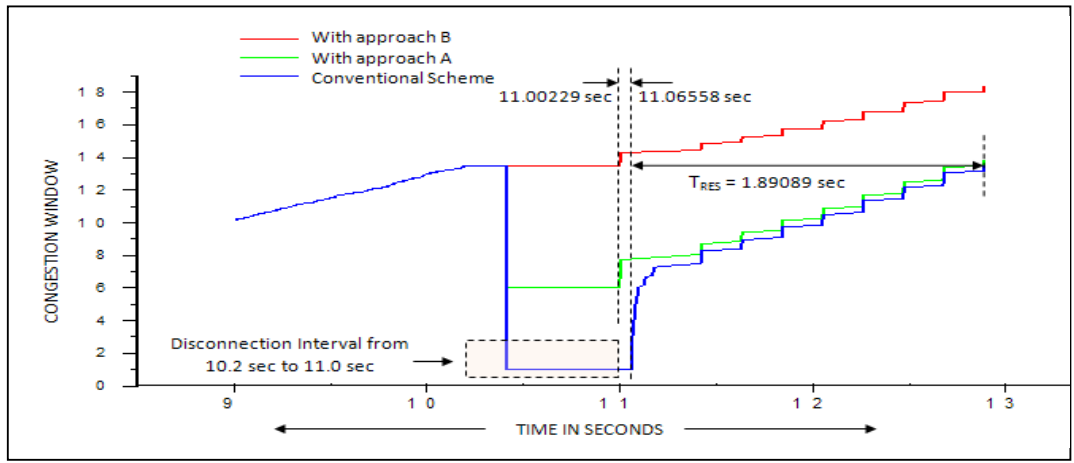

Figure. 10 Comparison in *cwnd* using different TCP and MAC implementations
($T_{LOSS}$ > RTO and short FTP with single disconnection)

Higher *cwnd* in presence of both of the approaches proposed in this paper enables the modified TCP to transmit data according to the network capacity and improves throughput performance. Figure 11 shows comparison between both the approaches in terms of the performance gain at TCP. Maximum of 23% of improvement above approach A is observed, for the loss duration corresponds to *RTT*. It may be noted that for prolonged loss durations, increase in $T_{CRD}$ may not be sufficient to trigger fast retransmit action at TCP, as the maximum value of *RL* is restricted to 24. As a result of this, for $T_{LOSS}$ >= *2.5 RTT*, no additional performance gain over approach A is observed with use of approach B.

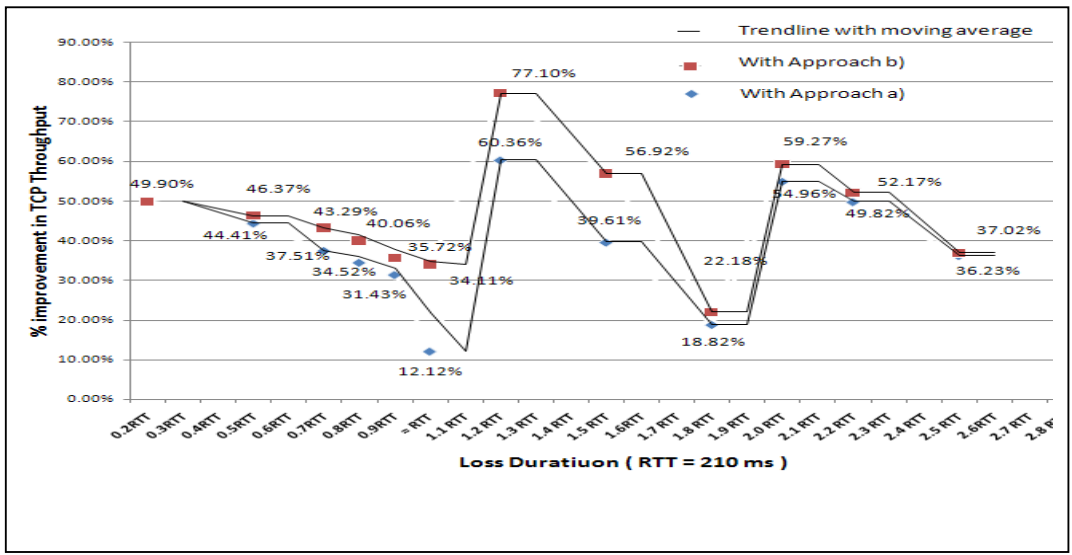

Figure. 11 Comparison in % improvement in TCP Throughput Vs $T_{LOSS}$

## 6. CONCLUSIONS

In view of the increasing demand of a wide range of services over Internet, it has become essential to augment the major protocol of Internet like TCP with abilities, which can combat the issues of the wireless networks with mobility. We observed that frequent link failures due to mobility lead to performance degradation resulting from the unnecessary serial timeouts coupled with congestion control. In this paper, we propose a pair of approaches to make TCP adaptable to wireless scenario and improve the end-to-end performance. Results of extensive simulations demonstrate significant performance improvement of TCP with the proposed modifications over the conventional TCP, when the network is suffering from frequent link failures. We summarize the following important observations made with reference to the proposed scheme.

1) Performance improvement using approach A is contributed to the fact that wireless link errors are not propagated to the sender TCP, due to extended loss recovery at link layer. Moreover, effect of increased *RL* with reduced data-rate is not suspected on parallel competing traffics.
2) In case of failure of MAC loss recovery using approach A, a new parameter *link_loss* introduced at MAC enables TCP to recognize link conditions and avoid conventional flow control accordingly. As congestion related losses are not visible at MAC, the modified scheme promises normal operations even on congested networks.
3) More than 45 % of performance improvement over conventional TCP is observed for a disconnection interval smaller than *RTT*. Performance improvement up to 70% is observed for the prolonged loss durations.

The current simulations are done for a single end to end connection with disconnections. Further work may involve testing performance of the scheme a) in presence of interference caused by neighbouring nodes b) in presence of BER.

**Authors**

Purvang Dalal is an Associate Professor at Dharmsinh Desai University Nadiad, Gujarat, INDIA. His research interests include cross-layer control for QoS in wireless networks, especially over 802.11 based WLAN. He is interested in the interoperation of Transport and Data Link layer controls and the fairness constraints due to packet loss and related loss recovery. He is having more than 10 years of experience in academics. He holds a M.E. Degree in Electronic and Communication Systems from Dharmsinh Desai University of Nadiad (2005), and a B.E. Degree from the Sardar Patel University, V.V.Nagar, Gujarat, (1998). He published 10 conference Papers. His other area of expertise includes CMOS VLSI Design and Analysis.

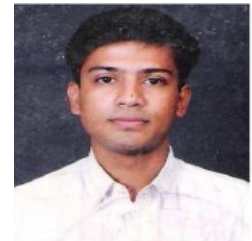

Nikhil J. Kothari received the degree in Electronics Engineering from Sardar Patel University, Vallabh Vidyanagar, Gujarat in 1985, the M.E. degree from the Gujarat University, Gujarat, in 1994, and the Ph.D. degree from Dharmsinh Desai University, Gujarat, in 2010. He joined the Department of Electronics & Communication Engineering, Dharmsinh Desai University, Nadiad, as a Lecturer in 1989, and has been a Professor since 2001. His current research interests are in wireless networks, satellite networks, and Internet Congestion Control.

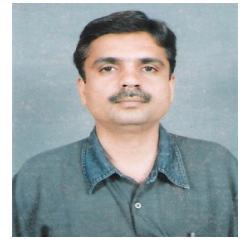

K. S. Dasgupta received his B.E. in Electronics & Telecommunications and his Master's degree in Computer Science from Jadavpur University in 1st class with Honours in 1972 and 1973 respectively. He obtained his Ph.D. in Electrical Engineering from Indian Institute of Technology Bombay in 1990. Dr. K S Dasgupta joined Space Applications Centre in 1974 and contributed significantly in the field of Image Processing and Satellite Communications, which honored him with a recognition of Outstanding/Distinguished Scientist. Before joining as Director of Indian Institute of Space Technology in December, 2010, he was the Deputy Director of SATCOM and Navigation Payload Area (SNPA) and Former Director-DECU (Development & Educational Communication Unit) of Indian Space Research Organization (ISRO). Dr. Dasgupta is senior member of IEEE, CSI and Fellow IETE. He has been awarded for excellent performance in year 2009 by ISRO.

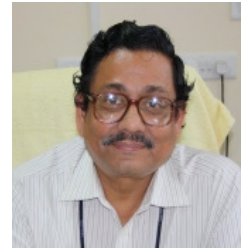